\documentclass[fleqn,twoside]{article}
\usepackage{espcrc2}
\usepackage{graphicx}

\title{The pion electromagnetic form factor}

\author{Frederic D. R. Bonnet$^{\rm a,b}$,
        Robert G. Edwards$^{\rm b}$,
        George T. Fleming$^{\rm b}$,
        Randy Lewis\address{Department of Physics, University of Regina,
                            Regina, SK, S4S 0A2, Canada},
        David G. Richards\address{Thomas Jefferson National Accelerator
                                  Facility, Newport News, VA 23606, USA} 
        }
\begin{document}

\begin{abstract}
A ratio of lattice correlation functions is identified from which the pion
form factor
can be obtained directly.  Preliminary results from quenched Wilson
simulations are presented.
\end{abstract}

\maketitle

\section{MOTIVATION}

The pion form factor is a convenient way to study the transition from
perturbative to nonperturbative QCD.  This transition is typically expected
to occur at a smaller $Q^2$ than for the nucleon, and should thus be more
easily attainable in experiments and in lattice studies.  Also, the asymptotic
normalization of the pion form factor is known in terms of $f_\pi$.
In addition, lattice simulations of the pion form factor benefit from the
complete absence of ``disconnected'' quark diagrams.\cite{Draper}

Some experimental data already exist, another experiment is in progress right
now at Jefferson Lab, and measurements at higher $Q^2$ are planned for the
future.\cite{Blok}.

Pioneering lattice QCD studies were carried out in the
1980's\cite{Draper,Woloshyn,Martinelli}, and only recently has the pion form
factor been revisited using lattice methods.\cite{vanderHeide,Nemoto}
In light of the substantial experimental effort, it is important to
produce lattice QCD results with significantly smaller quark masses than
have yet been employed,
and eventually to attain larger values of $Q^2$.  Unquenched simulations will
also be important.  Here, after a general discussion of correlation functions,
the issue of smaller quark masses will be addressed using quenched Wilson data.

\section{CORRELATION FUNCTIONS}

The electromagnetic form factor is obtained in lattice QCD simulations
by placing a charged pion creation operator at Euclidean time $t_i$, 
\begin{figure}[thb]
\includegraphics*[width=75mm]{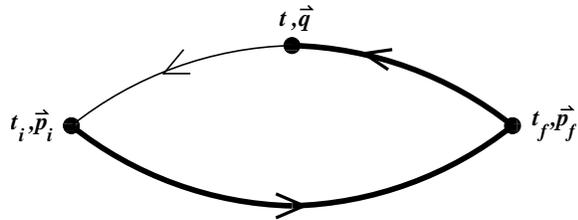}
\caption{The quark propagators used to compute the pion form
factor.}\label{threepoint}
\vspace{-5mm}
\end{figure}
a charged pion annihilation operator at $t_f$, and a vector current insertion
at $t$ as shown in Fig.~\ref{threepoint}.
A standard determination of 12 columns in the inverse of the quark matrix
provides the two propagators that originate from $t_i$.
The remaining propagator is obtained for any particular annihilation
operator, having definite momentum $\vec{p}_f$, by a two step procedure:
the propagator from $t_i$ to $t_f$ is multiplied by the annihilation
operator and that product is used as the input for a second 12-column linear
algebra step.
This determines the entire thick quark line of Fig.~\ref{threepoint}.
The momenta at the source ($\vec{p}_i$) and at the insertion ($\vec{q}$)
can then be varied freely without computing new propagators.
Since the largest $Q^2$ is attained in the Breit frame
($\vec{p}_f = -\vec{p}_i$), the simple choice of $\vec{p}_f=\vec{0}$ is not
always desirable.

The form factor, $F(Q^2)$, is defined by
\begin{eqnarray}
&&\!\!\!\!\!\!\!\!\!\!\!\!
\left<\pi(\vec{p}_f)\left|V_\mu(0)\right|\pi(\vec{p}_i)\right>_{\rm continuum}
 \\
&&\!\!\! = Z_V\left<\pi(\vec{p}_f)\left|V_\mu(0)\right|\pi(\vec{p}_i)\right>
= F(Q^2)(p_i+p_f)_\mu \nonumber
\end{eqnarray}
where $V_\mu(x)$ is the chosen vector current.
$F(Q^2)$ can be extracted from ratios of lattice correlation functions.
The three-point correlator is
\begin{eqnarray}
&&\!\!\!\!\!\!\!\!\!\!\!\!
\Gamma_{\pi\mu\pi}^{AB}(t_i,t,t_f,\vec{p}_i,\vec{p}_f)
 = a^9\sum_{\vec{x}_i,\vec{x}_f}
e^{-i(\vec{x}_f-\vec{x})\cdot\vec{p}_f}
\nonumber \\
&&\times e^{-i(\vec{x}-\vec{x}_i)\cdot\vec{p}_i}
   \left<0\left|\phi_B(x_f)V_\mu(x)\phi_A^\dagger(x_i)\right|0\right>
\end{eqnarray}
where $A\in(L,S)$ and $B\in(L,S)$ denote either ``local'' or ``smeared''.
Inserting complete sets of hadron states and requiring $t_i\ll t\ll t_f$,
gives
\begin{eqnarray}
&&\!\!\!\!\!\!\!\!\!\!\!\!
\Gamma_{\pi\mu\pi}^{AB}(t_i,t,t_f,\vec{p}_i,\vec{p}_f) \to
     \left<0\left|\phi_B(x)\right|\pi(\vec{p}_f)\right>
\nonumber \\
&&
     \times\left<\pi(\vec{p}_f)\left|V_\mu(x)\right|\pi(\vec{p}_i)\right>
     \left<\pi(\vec{p}_i)\left|\phi_A^\dagger(x)\right|0\right> \nonumber \\
&&
     \times\frac{a^3}{4E_fE_i} e^{-(t_f-t)E_f}e^{-(t-t_i)E_i}.
\end{eqnarray}
Similarly for the two-point correlator,
\begin{eqnarray}
&&\!\!\!\!\!\!\!\!\!\!\!\!
\Gamma_{\pi\pi}^{AB}(t_i,t_f,\vec{p})
 \to \left<0\left|\phi_B(x_i)\right|\pi(\vec{p})\right>
\nonumber \\ && \times
     \left<\pi(\vec{p})\left|\phi_A^\dagger(x_i)\right|0\right>
     \frac{a^3}{2E}e^{-(t_f-t_i)E}.
\end{eqnarray}
Finally, consider operators that are temporally local, but spatially smeared
democratically among the three spatial axes:
\begin{eqnarray}
a^2\left<0\left|\phi_L(x)\right|\pi(\vec{p})\right>
 &=& Z_Le^{ix\cdot p}, \\
a^2\left<0\left|\phi_S(x)\right|\pi(\vec{p})\right>
 &=& Z_S(|\vec{p}|)e^{ix\cdot p}.
\end{eqnarray}
{}From this discussion, it is found that the following expression for the form
factor is independent of $Z_L$, $Z_S(|\vec{p}|)$, and all time exponentials:
\begin{eqnarray}
F(Q^2) &=& \frac{\Gamma_{\pi 4\pi}^{AB}(t_i,t,t_f,\vec{p}_i,\vec{p}_f)
   \Gamma_{\pi\pi}^{CL}(t_i,t,\vec{p}_f)}
  {\Gamma_{\pi\pi}^{AL}(t_i,t,\vec{p}_i)
   \Gamma_{\pi\pi}^{CB}(t_i,t_f,\vec{p}_f)}
\nonumber \\ && \times
   \left(\frac{2Z_VE_f}{E_i+E_f}\right), \label{theratio}
\end{eqnarray}
where the indices $A$, $B$ and $C$ can be either $L$ (local)
or $S$ (smeared).
For a conserved current, $Z_V\equiv1$.

\section{ACTION AND PARAMETERS}

Our initial explorations have used the (quenched) Wilson action with $\beta=6$,
so $a\sim0.10$ fm.
Periodic boundary conditions are used except for the quark temporal
boundaries, which are Dirichlet.  Hopping parameters and lattice sizes are
shown in Table~\ref{masstable}; the resulting pseudoscalar and vector meson
masses are also listed.  Notice that the smallest pion mass is
less than 300 MeV, and therefore significantly below those used in previous
studies of the pion form factor.
The pion operators are fixed at $t_i=7$ and $t_f=21$ except for the
$32^3\times48$ lattices, where $t_f=28$.
The standard point-split conserved vector current is used.
A gauge-covariant Gaussian smearing is employed at the source,
\begin{equation}
b(x) \to \left(1+\frac{\omega\vec\nabla U}{N}\right)^Nb(x).
\end{equation}
The various results shown here are based on the analysis of
between 80 and 155 configurations.
\begin{table}
\caption{The chosen hopping parameters and numbers of lattice sites, along with
the computed pseudoscalar and vector meson masses.}\label{masstable}
\begin{tabular}{llll}
$\kappa$ & lattice & $am_{PS}$ & $am_V$ \\
\hline
0.1480 & $16^3\times32$ & 0.673 & 0.712 \\
0.1520 & $16^3\times32$ & 0.477 & 0.549 \\
0.1540 & $16^3\times32$ & 0.364 & 0.468 \\
0.1555 & $24^3\times32$ & 0.259 & 0.398 \\
0.1563 & $24^3\times32$ & 0.179 & 0.358 \\
0.1566 & $24^3\times32$ \& $32^3\times48$ & 0.145 & 0.343
\end{tabular}
\vspace{-5mm}
\end{table}

\section{PRELIMINARY RESULTS}

\begin{figure*}[thb]
\begin{center}
\includegraphics*[height=59mm]{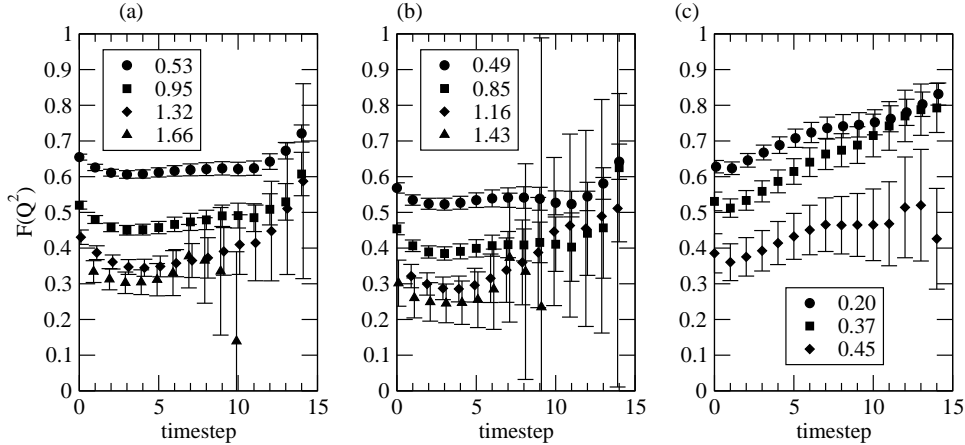}
\end{center}
\vspace{-13mm}
\caption{Pion form factor data versus timestep, for (a) $\kappa$=0.1520,
(b) $\kappa$=0.1540 and (c) $\kappa$=0.1563.  Numerical values of $Q^2$ in GeV
are shown in the legends.}\label{plateaus}
\vspace{-2mm}
\end{figure*}

Representative plots of the data are shown in Fig.~\ref{plateaus}.
For each of the smaller hopping parameters, there is a reasonable plateau at
timesteps sufficiently far from the pion operators (timesteps 0 and 14).
For $\kappa=0.1563$ and 0.1566, the possible plateau spans only a few
timesteps.

\begin{figure*}[hbt]
\begin{center}
\includegraphics*[height=59mm]{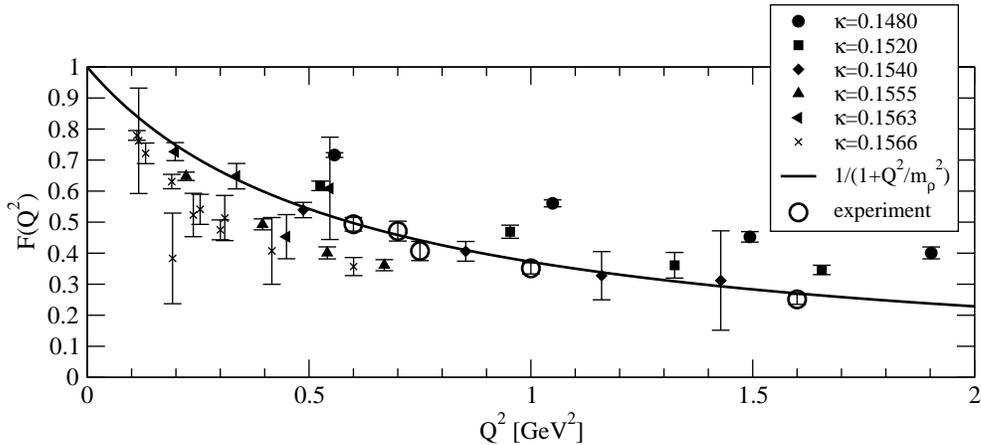}
\end{center}
\vspace{-13mm}
\caption{Results for the pion form factor as a function of $Q^2$ for each
of the available $\kappa$ values.  Experimental
measurements\protect\cite{Volmer} and the monopole
approximation are also shown.}\label{finalplot}
\vspace{-2mm}
\end{figure*}

Correlated fits to the data have produced Fig.~\ref{finalplot} which shows
the form factor as a function of $Q^2$ for each available $\kappa$.
For comparison, the monopole curve $1/(1+Q^2/m_\rho^2)$ is also shown
where $m_\rho$ denotes the physical $\rho$ meson mass.
Notice that the experimental measurements from Ref.~\cite{Volmer} lie on the
monopole curve.
For each $\kappa$ the lattice results also display a monopole form,
and $F(Q^2)$ is found to decrease with decreasing quark mass as expected
since the effective vector meson mass of the monopole form also shrinks
with decreasing quark mass.

In conclusion, Eq.~(\ref{theratio}) is found to be a useful method for
extracting the pion form factor.  Simulations with
domain-wall fermions are underway.

This work was supported in part by the Natural Sciences and Engineering
Research Council of Canada and by the U.S. Department of Energy under
contract DE-AC05-84ER40150.  Computations were performed on the 128-node
Pentium IV cluster at JLab.

\end{document}